\input{aipcheck}

\documentclass[
    ,final            % use final for the camera ready runs
  ]
  {aipproc}

\layoutstyle{6x9}

\usepackage{epsfig}
\usepackage[fleqn,sumlimits,intlimits]{amsmath}
\usepackage{latexsym,amssymb}
\usepackage[english]{babel}
\usepackage{verbatim}
\usepackage{color}
\usepackage{varioref}

\newcommand{\bea}{\begin{eqnarray}}
\newcommand{\eea}{\end{eqnarray}}
\newcommand{\beq}{\begin{equation}}
\newcommand{\eeq}{\end{equation}}

\newcommand{\bal}{\begin{align}}
\newcommand{\eal}{\end{align}}
\newcommand{\bit}{\begin{itemize}}
\newcommand{\eit}{\end{itemize}}

\newcommand{\half}{{\frac{1}{2}}}

\newcommand{\C}[1]{{\mathcal{#1}}}

\begin{document}

\title{CDT coupled to dimer matter:\\ An analytical approach via tree bijections}

\classification{04.60.Nc,04.60.Kz,04.60.Gw}
\keywords      {Causal dynamical triangulations (CDT), tree bijections, hard dimer model.}

\author{Max R. Atkin}{
  address={Fakult\"{a}t f\"{u}r Physik, Universit\"{a}t Bielefeld, Postfach 100131, Bielefeld, Germany},
  email={matkin@physik.uni-bielefeld.de}
}

\author{Stefan Zohren}{
  address={Department of Physics, 
  %Pontifica Universidade Cat\'olica do Rio de Janeiro, 
  PUC-Rio, R. Marqu\^es de S\~ao Vicente 225, Rio de Janeiro, Brazil},
 altaddress={Rudolf Peierls Centre for Theoretical Physics, 1 Keble Road, Oxford, UK} ,% additional visiting address
email={zohren@fis.puc-rio.br}}

\begin{abstract}
We review a recently obtained analytical solution of a restricted so-called hard dimers model coupled to two-dimensional CDT. The combinatorial solution is obtained via bijections of causal triangulations with dimers and decorated trees. We show that the scaling limit of this model can also be obtained from a multi-critical point of the transfer matrix for dynamical triangulations of triangles and squares when one disallows for spatial topology changes to occur. 
\end{abstract}

\maketitle

%%%%%%%%%%%%%%%%%%%%%%%%%%%%%%%%%%%%%%%%%%%%
%% MAINMATTER
%%%%%%%%%%%%%%%%%%%%%%%%%%%%%%%%%%%%%%%%%%%%

\section{CDT and matter}
\begin{figure}[t]
\centering 
\includegraphics[scale=0.48]{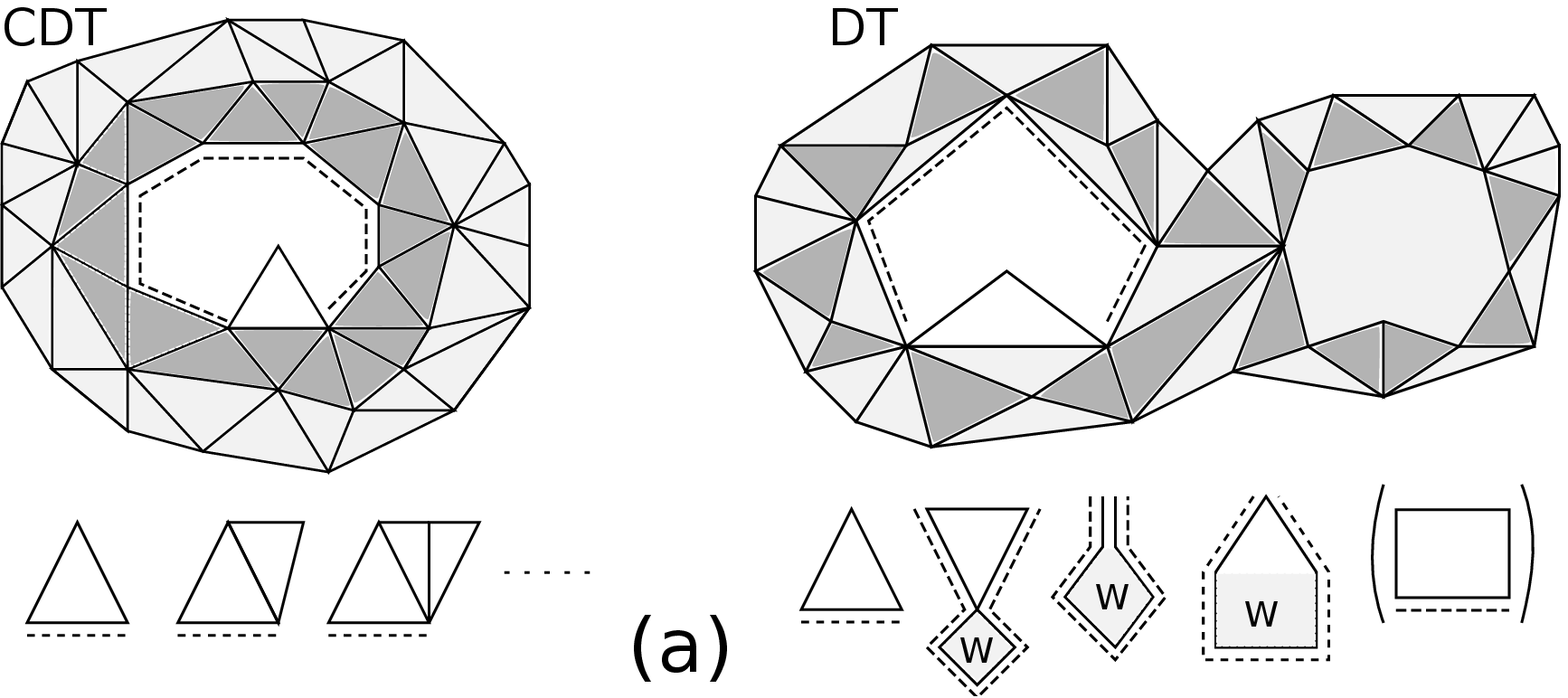}
%\label{CDTandDT}
\qquad 
\includegraphics[scale=0.35]{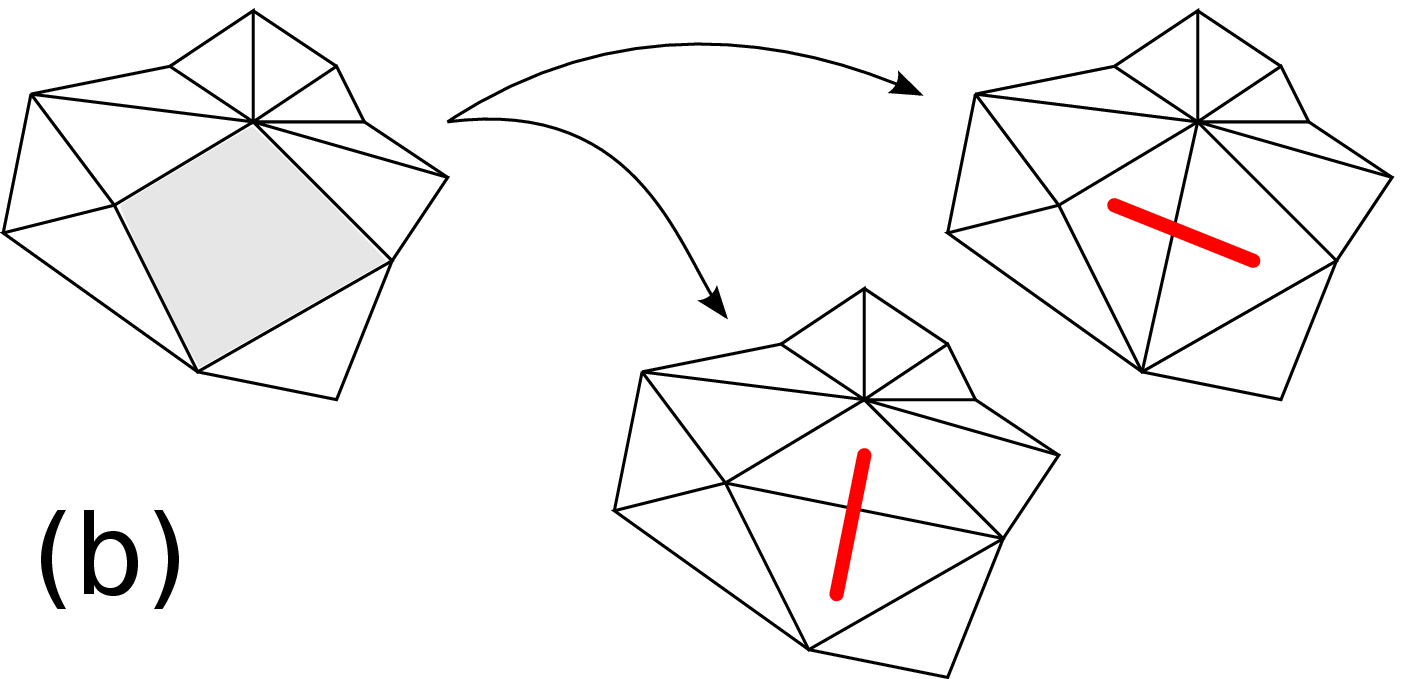}
\caption{{\bf (a)} Here we show an example of a two-dimensional CDT and DT. The exit boundary is the one marked by a dotted line. The pieces beneath each diagram represent the pieces that can be added to the exit boundary to construct the transfer matrix, there $W$ represents a disc function. The final square piece in brackets does not appear in the pure case. Here we have coloured the triangulations by light and dark grey to represent alternate applications of the transfer matrix. {\bf (b)} The mapping between squares and hard dimers. This mapping does not respect the causal structure of CDT.}
\label{squareTodimers} 
\end{figure}

The idea of Causal Dynamical Triangulation (CDT) \cite{CDT} follows from a long tradition of discrete approaches to quantum gravity in which the path integral over geometries is regulated by approximating it by a sum over triangulations composed of flat triangles or more generally polygons. The regularisation is removed at the end of a calculation by finding an appropriate scaling limit. In the case of Dynamical Triangulation (DT) \cite{DT} and CDT the triangles have fixed length sides of size equal to the cut-off and all geometry is encoded in how they are glued together. In particular curvature is localised at the points where multiple triangles meet. This means the sum over triangulations directly implements a sum over discrete geometries in contrast to summing over metrics, in which one must mod out the group of diffeomorphisms. The difference between DT and CDT is in the class of triangulations summed over, with CDT introducing an extra causal constraint corresponding to a preferred time-slicing with respect to which no spatial topology change may occur. This difference between two-dimensional CDT and DT, which are unrestricted planar triangulations, is shown in Figure \ref{squareTodimers} (a).

Much work has been done investigating the properties of pure (i.e.\ withour matter coupling) quantum gravity defined through CDT, in $D=2,3,4$ dimensions (see \cite{review} for a recent review): For $D>2$, work has been numerical in nature and produced impressive results demonstrating the existence of a phase in which an extended de-Sitter space-time exists. There have also been interesting results related to the UV properties of the theory with indications that in the UV limit $D=2$. In two dimensions analytic calculations are possible and the theory has been reformulated as a matrix model \cite{matrix} showing that the $D=2$ can be interpreted as a string worldsheet theory. The simplification for $D=2$ is due to the topological nature of the Riemann-Hilbert action, which leads to an action that only depends on the space-time volume weighted by the cosmological constant $\Lambda$. The DT and CDT programs therefore replace the gravitational path integral by, 
\beq
\hat{\C{Z}}(\Lambda,\Gamma) = \int d\C{T} \int \C{D} g e^{-\Lambda A_g + S_{g,\C{T}}(\Gamma)} %= \int d\C{T} \hat{Z}(\C{T};\Lambda,\Gamma)
%\nn\\ 
 \rightarrow \C{Z}(\lambda,\gamma) = \sum_\tau \sum_T \frac{1}{C_T} e^{- \lambda n_T + s_{T,\tau}(\gamma ) }  %= \sum_\tau Z(z,\tau),
\eeq
where %we have denoted continuum quantities by hats, 
$g$ is the space-time metric, $A_g$ is the continuum volume of the space-time, $\C{T}$ represents the matter fields and $ S_{g,\C{T}}(\Gamma)$ the action of the matter coupling which depends on a set of coupling constants labelled $\Gamma$. On the discrete side, the integral over metrics is replaced by a sum over triangulations $T$ (including a symmetry factor $1/C_T$) and the integral over continous matter configurations $\C{T}$ by a sum over discrete matter configurations $\tau$. Further, $A_g$ is replaced by the number of triangles $n_T$ of $T$ and the coupling constants $\lambda$ and $\gamma$ are ``bare'' analogues of $\Lambda$ and $\Gamma$. 

Matter coupled to CDT has been studied numerically in \cite{matter-num} with indications that the back-reaction on the geometry is not as severe as in DT. This has prompted suggestions that conformal matter with central charge $c>1$ may be able to couple to CDT \cite{breaking}. Some analytic work has also been done \cite{DiF} in which new scaling limits were found however this work had the disadvantage that it was unclear what conformal field theory should describe the matter sector in the continuum limit. 

A simple matter model coupled to CDT which can be studied analytically are so-called hard dimers (see \cite{Staudacher} for dimers coupled to DT). In this model dimers with a fugacity $\xi$ can join two adjacent triangles and they are hard in the sense that a single triangle can only contain a maximum of one dimer. The analytical derivation of the partition function reduces to a combinatorial problem of finding the number $\C{N}(n;k)$ of causal triangulations (of a sphere) with $n$ triangles with $k$ dimers, since
\beq
\C{Z}(z,\xi) = \sum_{k=0}^\infty \sum^\infty_{n = 0} \C{N}(n;k) z^n \xi^k
\eeq
where $z=e^{- \lambda}$ is the fugacity of a triangle while $\xi$ is the fugacity of a dimer as introduced above. From the point of view of statistical mechanics, $\C{N}(n;k)$ can be thought of as the micro-canonical ensemble while $\C{Z}(z,\xi)$ is the canonical ensemble. 

In what follows we review some recent results by the authors \cite{dimer} regarding a combinatorial solution for determining $\C{N}(n;k)$ for a restricted class of hard dimers coupled to two-dimensional CDT. Simultaneous work by Ambj{\o}rn et al.\ \cite{multi} shows that the matrix model formulation of pure two-dimensional CDT \cite{matrix} can been extended to a higher multi-critical point in agreement with the scaling limit of the combinatorial solution, thus providing evidence that both these approaches yield the correct continuum limit of hard dimers coupled to two-dimensional CDT. Before reviewing the combinatorial solution obtained in \cite{multi} we present a new alternative approach to the problem using a scaling limit of a transfer matrix for CDT with squares reminiscent of the higher multi-critical point in the matrix model \cite{multi} and the related peeling procedure \cite{higher}.

\section{A transfer matrix for CDT with squares}
The partition function of two-dimensional pure CDT on a cylinder was first obtained using transfer matrix techniques \cite{CDT}. This amounts to determining the quantity $\C{N}(n,l_i,l_f,t)$, i.e. the number of triangulations of a cylinder with $n$ triangles, where we have added the boundary condition that the entrance and exit loops of length $l_i$ and $l_f$ are separated by a distance/time $t$. The derivation in \cite{CDT} made explicit use of the preferred time-slicing by identifying the edges at a constant distance from the entrance loop, or equivalently constant time, with the subsystem acted on by the transfer matrix. A similar computation in DT was performed earlier \cite{DT}; here however there is no time-slicing, instead the transfer matrix acts on subsystems defined as the triangles whose vertex in the dual graph is at constant dual-distance from the entrance loop. This difference is illustrated in Figure \ref{squareTodimers} (a). It is important to note that one has a composition law,
\beq
Z(z,l_i,l_f, t+1) = \sum^\infty_{l=0} Z(z,l_i,l,t) Z(z,l,l_f,1).
\eeq
which allows one to identify the transition matrix as $Z(z,l_i,l_f,1)$. Technically it is easier to work with the generating functions $\C{Z}(z, x, y,t)$ for these quantities. We then have the composition law,
\beq
\label{Zcomp}
\C{Z}(z,x,y, t+1) = \oint \frac{dw}{2\pi i w}  \C{Z}(z,x,w^{-1},t)\C{Z}(z,w,y,1).
\eeq
To compute the transfer matrix $\C{Z}(z,w,y,1)$ we must construct the generating function for all possible ways one can obtain an exit loop from a given entrance loop in one step. We show in Figure \ref{squareTodimers} (a) the possible pieces of triangulation from which the DT transfer matrix can be built. It is important to note that in DT the effect of a change in the topology of a slice is contained entirely in the local moves in Figure \ref{squareTodimers} (a).

Analogous to the construction in \cite{matrix} and \cite{baby} it is suggestive to propose that {\emph{forbidding spatial topology changes with respect to the spatial slices in the DT transfer matrix construction produces triangulations in the same universality class as CDT}}. Assuming this proposal for a moment, we may then go further and add squares to the transfer matrix as was done in \cite{DTtrans}. The intention of \cite{DTtrans} was to study DT coupled to hard dimer matter; here, by forbidding topology change we expect to recover CDT coupled to some form of matter. Explicitly we have,
\bea
\C{Z}(z,x,y,1) \!=\! \left( 2xy^2z + x^2yz + 3 z^2\xi xy^3 \right) \left(1\! +\! K\! +\!K^2\! +\!\ldots \right)
\!=\! \frac{2xy^2z + x^2yz + 3 z^2\xi xy^3}{1 - K}
\eea
where the first factor adds a single marked triangle (see \cite{DTtrans} for details) and $K = z xy^2 + z x^2y + z^2\xi x y^3$, represents the possibility of adding the first, second or fifth piece under the DT diagram in Figure \ref{squareTodimers} (a). Note that we have set $W = 1$ in each of these pieces to prevent topology change. Finally, by setting $\xi = 0$ we can return to a pure triangulation. To remove the regularisation we look for a scaling limit by making the ansatz $x = x_c e^{-a X}$, $y = x_c^{-1} e^{-a Y}$ and $z = z_c e^{-a^k \Lambda}$ where $k$ is a constant and letting $a \rightarrow 0$. To find the critical point $(x_c,z_c)$ we demand that to leading order, $\C{Z}(z,x,y,1) = a^{-1}(X + Y)^{-1} + O(1)$; this is equivalent to requiring $\C{N}(n,l_i,l_f,0) = \delta_{l_i,l_f}$. This requirement gives $z_c$ as a function of $\xi$. By substituting the scaling ansatz into \eqref{Zcomp} we obtain in general,
\beq
\hat{\C{Z}}(\Lambda,X,Y, t+1) = \int^{i \infty}_{-i \infty} \frac{a d\zeta}{2\pi i} \hat{\C{Z}}(\Lambda,X,-\zeta,t) \left[\frac{1}{a(Y + \zeta)} + \ldots + a^{\eta-1} \frac{F(Y,\zeta,\Lambda)}{(Y + \zeta)^2} \right].
\eeq
where the expression in the square brackets corresponds to $\C{Z}(\Lambda,X,\zeta,1)$, $\eta$ is a function of $k$, $F$ is some unspecified function and the dots represent terms which contain no poles in $\zeta$. By closing the contour and applying the residue theorem we obtain a differential equation of the form,
\beq
\label{PropH}
\partial_T \hat{\C{Z}}(\Lambda,X,Y, T) = \partial_{Y} \left[ G_F(\Lambda,Y)  \hat{\C{Z}}(\Lambda,X,Y, T) \right]
\eeq
where $G_F$ is a function derived from $F$ and $T = a^{\eta} t$ is a continuum distance/time. We now present two distinct scaling limits;
\begin{itemize}
\item For $\xi = 0$; $x_c = 1$ and $z_c = 1/2$. We also have $\eta = 1$ with $G_F = Y^2 - \Lambda$. This is precisely the result derived in \cite{CDT,matrix} for pure CDT. This justifies our earlier proposal.
\item For $\xi \neq 0$; we find we must also set $\xi$ to a critical value of $ \xi_c = -\frac{1}{3}$ to obtain a new scaling limit. We then have $x_c = z_c = 3^{-\frac{1}{2}}$. We also have $\eta = 2$ with $G_F = Y^3 - \Lambda$. We can also extract a scaling exponent from $d\log z/d\xi \sim (\xi - \xi_c)^\sigma$, finding $\sigma = \half$. This is precisely the result derived in \cite{multi,dimer} for multi-critical CDT. Please consult \cite{higher} for a discussion of the correct boundary condition to impose on \eqref{PropH}.
\end{itemize}
There are two important observations to make about the $\xi \neq 0$ case. Firstly, we have only allowed squares to be added by attaching them along one edge. In the DT case \cite{DTtrans} squares were added in every manner; if we were to do this here we would not find a new scaling limit. Secondly, the scaling dimension of $z$ and $t$ differs from the $\xi = 0$ case, implying that the geometry of the multicritical phase is in some sense fractal. Both of these properties show that the multicritical phase is more delicate in CDT than DT. To address why this is the case, we recall why we expect the addition of squares to lead to a scaling limit corresponding to the addition of hard dimers.
A hard dimer model adds objects called hard dimers to a triangulation, with the property that they occupy two adjacent triangles and no two dimers can occupy the same triangle. The mapping between dimers and squares is shown in Figure \ref{squareTodimers} (b); note that in CDT the two possible dimers corresponding to the same square do not always respect the causal structure and hence it is clear this mapping breaks down.

\begin{figure}[t]
\centering 
\includegraphics[scale=0.4]{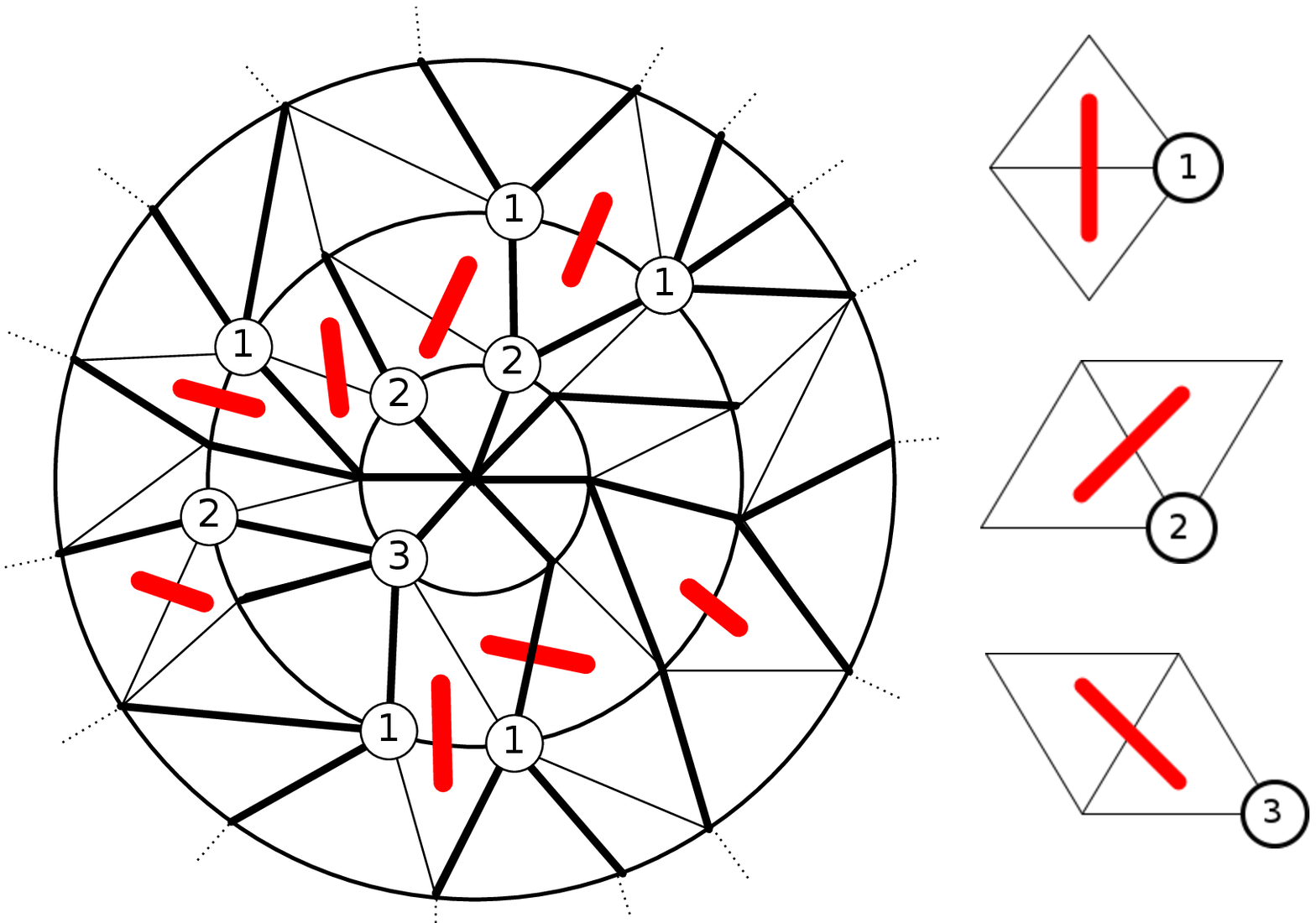}
%\label{CDTbijection1}
\quad 
\includegraphics[scale=0.52]{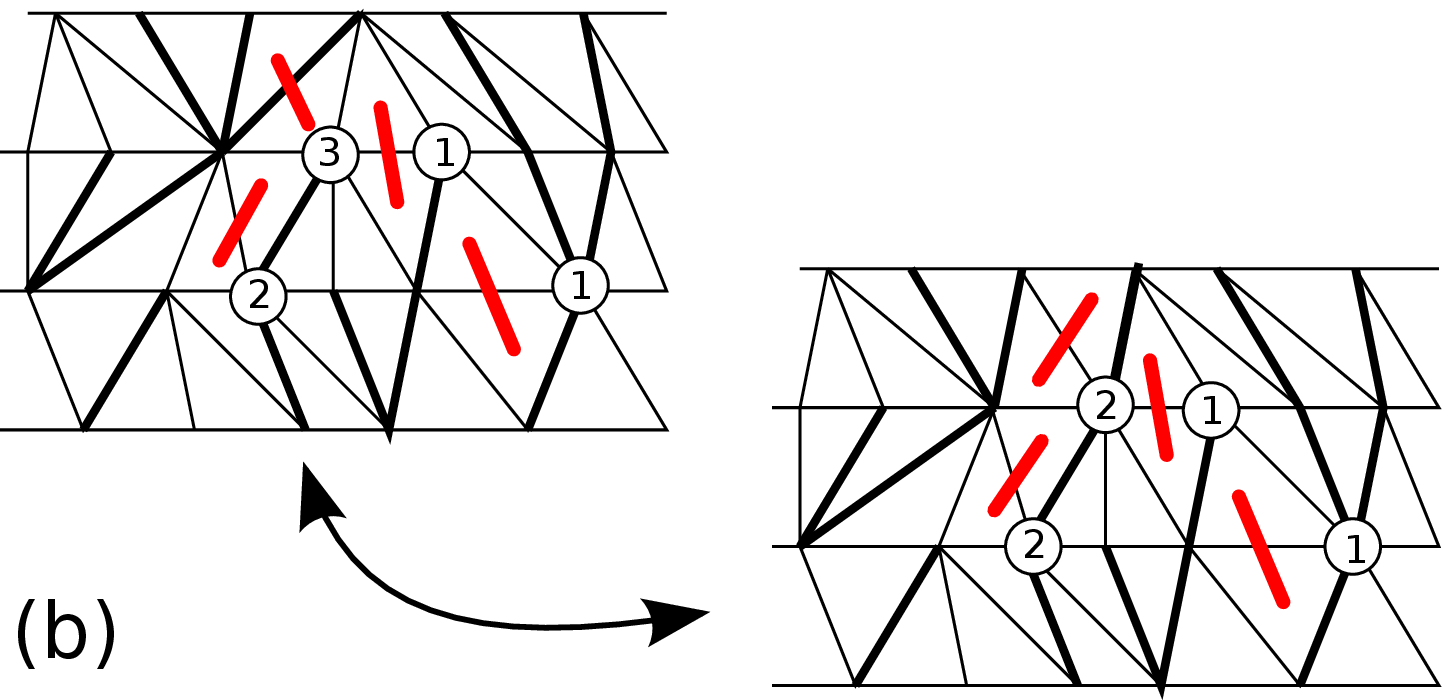}
\caption{{\bf (a)} A bijection between hard dimers in CDT and labelled trees. The tree is shown in thickened lines and the labels are added using the rules on the right. {\bf (b)} An example of the bijection between distinct trees with type-3 dimers and those without.}
\label{CDTbijection} 
\end{figure}

\section{CDT with restricted hard dimers}

We now review the results of \cite{dimer} and argue the above model does indeed correspond to hard dimers on CDT. We compute $\C{N}(n,k;\tau)$, where $k$ is the number of dimers via the bijection to labelled tree graphs shown in Figure \ref{CDTbijection} (a). Bijections between (C)DT and trees have been known for some time (see \cite{treesDT} for DT and \cite{treesCDT} for CDT) so it is no surprise that dimers can by adding by labelling the tree. The difficulty lies in counting such trees since the bijection removes links and hence local interactions in the CDT can become non-local in the tree, thereby preventing the trees being defined recursively and hence enumerated.

In \cite{dimer} we strike a balance; we include enough non-local interactions to see new critical behaviour but not so much that the calculation is impossible. This can be seen in Figure \ref{CDTbijection} (a); we only allow a restricted class of dimers to appear. Note that the type-3 dimers are non-local, however they can be accounted for by noting there exists a bijection between trees containing such dimers and distinct trees in which they are replace by type-2 dimers. This bijection is shown in Figure \ref{CDTbijection} (b). Dimers of type one and two only affect dimers which are adjacent to them in the tree. Hence trees containing only these can be counted recursively. Hence we obtain a recursive formula for the generating function $G_d(z,\xi)$ of $\C{N}_d(n,k)$, where $\C{N}_d(n,k)$ is the number of trees with $n$ vertices, $k$ dimers and whose root vertex is labelled by $d$;
\beq
G_0(z,\xi) = \frac{1}{1-z^2(G_0 + \xi G_1 + 2\xi G_2 )}  \qquad \mathrm{and} \qquad G_2(z,\xi) = z^2(G_0 + 2\xi G_2)G_0.
\eeq
where $d=0$ implies no label, and $G_1 = G_0$. Note the factor of two with each $G_2$; this accounts for the type-3 dimers. Solving for $G_0$ we obtain the partition function $\C{Z}(z,\xi) = G_0$ for dimers coupled to CDT on the sphere. From this we can compute $\sigma = \half$. Furthermore we find in the scaling limit with $\xi = \xi_\mathrm{critical}$ that $G_0 - G_\mathrm{critical} \sim \Lambda^{1/3}$. This agrees with the computations in \cite{multi} and with those above.

\section{Conclusions}

We reviewed a recent work by the authors \cite{dimer} in which a combinatorial solution of a restricted class of hard dimers coupled to two-dimensional CDT was obtained. The scaling limit of this model arguably lies in the same universality class as the full model (i.e.\ unrestricted class) of hard dimers coupled to two-dimensional CDT. This continuum model can also be obtained from a higher multi-critical point of the CDT matrix model as was shown by Ambj{\o}rn et al. \cite{multi}. Inspired by the derivation of the higher multi-critical point in the CDT matrix model \cite{multi} and the corresponding peeling procedure \cite{higher}, we proposed an analogous solution through a higher multi-critical point of the DT transfer matrix with spatial topology changes suppressed. This provides us with a different and interesting perspective on the problem.

%%%%%%%%%%%%%%%%%%%%%%%%%%%%%%%%%%%%%%%%%%%%%%%%
%% BACKMATTER
%%%%%%%%%%%%%%%%%%%%%%%%%%%%%%%%%%%%%%%%%%%%%%%%

%\begin{theacknowledgments}
This work has been supported by the U.\ of Bielefeld and STFC grant ST/G000492/1.
%\end{theacknowledgments}



\bibliographystyle{aipproc}   % if natbib is available

%\bibliography{random_graphs,spectral_dimension,cdt_dt_mm}

 %{\typeout{}
  %\typeout{******************************************}
  %\typeout{** Please run "bibtex \jobname" to optain}
 %\typeout{** the bibliography and then re-run LaTeX}
 %\typeout{** twice to fix the references!}
%\typeout{******************************************}
 %\typeout{}
 %}

\end{document}